\documentclass[aps, prd, nofootinbib]{revtex4-2}
\usepackage{aas_macros}
\usepackage{graphicx}
\usepackage{amsfonts, amsmath, mathrsfs}
\usepackage{xcolor}
\usepackage{hyperref}
\usepackage[final]{changes}
\begin{document}

\title{Measuring type Ia supernova angular-diameter distances with intensity interferometry}

\author{A.\ G.\ Kim}
\email{agkim@lbl.gov}
\author{P.\ E.\ Nugent}
\affiliation{Lawrence Berkeley National Laboratory\\
Berkeley, California 94720, USA}
\author{Xingzhuo Chen}
\affiliation{Texas A\&M Institute of Data Science Texas A\&M University 3156 TAMU,\\ College Station, Texas, 77843-3156, USA}
\author{L.\ Wang}
\affiliation{Department of Physics and Astronomy 4242 TAMU\\
College Station, Texas 77843-4242, USA}
\author{J. T. O'Brien} % 0000-0003-3615-9593
\affiliation{Department of Astronomy University of Illinois Urbana-Champaign\\
Champaign, Illinois 61801-3633, USA}
\affiliation{TARDIS Collaboration}

\begin{abstract}
This paper investigates the potential of intensity interferometry, based on the Hanbury Brown-Twiss effect, for measuring supernova sizes and distances.
With optimized telescope positioning, observing strategy, and advancements in single-photon detection technology, this method can provide precise angular size measurements of supernovae with apparent magnitudes as bright as 12~mag. For type~Ia supernovae, this limiting brightness corresponds to a local volume extending to redshift $z \sim 0.004$ and an anticipated discovery rate of approximately 1 event per year.
% Through optimized telescope positioning, observing strategy,
% and advancements in single-photon detection technology,  this method can provide precise angular size measurements of type Ia supernovae 
% as bright as 12~mag, corresponding to a local volume out to $z\sim0.004$,
% with an anticipated rate of $\sim 1$ events per year.
The combination of angular size data with known physical dimensions enables accurate distance determination. 
Multiple telescope pairs at different relative positions allow tomographic mapping of the ejecta structure while reducing distance uncertainties.
As type Ia supernovae serve as standardizable candles for measuring the Universe's expansion history, combining intensity interferometry distances with the supernova Hubble diagram facilitates measurements of the Hubble constant 
$H_0$.
\end{abstract}
\maketitle

\section{Introduction}

Intensity interferometry, founded on the Hanbury Brown-Twiss effect \cite{1954PMag...45..663B, 
1956Natur.178.1046H}
(known as the GGLP effect
in particle physics \cite{Goldhaber:1960sf}), 
has gained renewed interest with 
successful measurements coming from the
Cherenkov telescopes VERITAS,
MAGIC, and H.E.S.S.\ 
\cite{2020MNRAS.491.1540A,10.1093/mnras/staa3607, 2024MNRAS.52712243Z, abe2024performance,  2024ApJ...966...28A, 10.1093/mnras/stae2643}
and deployments at 1m to 2m class telescopes
\cite{2021MNRAS.506.1585Z,10.1093/mnras/staa3607,2023AJ....165..117M}.
The success is largely due to
the emergence of fast single-photon detector arrays
using either
single-photon avalanche detectors (SPAD)
\cite{10.1117/12.2253598, 2016Senso..16..745P,2018JaJAP..57j02A3L} or superconducting nanowire single-photon detectors (SNSPD) \cite{divochiy2008superconducting, zhu2020resolving, korzh2020demonstration,10.1093/mnras/staa3607}.
These technologies facilitate multiplexed, precise time-correlation measurements across telescopes
\cite{2023OExpr..3144246C,2024arXiv240613959F,2024arXiv241108417T}. 
These advancements enable new scientific reach for  the study of resolved sources and the
field of astronomical astrometry \cite{2022OJAp....5E..16S}.

The primary observables in intensity interferometry are the correlations
of photon streams at different positions.
The coherence of optical waves at different
coordinates and times can be decomposed
into baseline (physical separation) and time-lag contributions.
The time-lag contribution  is what gives rise to the Hanbury Brown-Twiss effect.
The  light detected from astronomical sources have coherent temporal fluctuations
that are directly
related to the transmitted spectrum.
Two telescopes observing the same source over a common time period and passband measure the same fluctuations, which gives rise to a
coherence in their time-dependent signals.
The baseline contribution to coherence depends on the interference pattern produced by \deleted{the scene,} the
light emission as a function of angular position.
This pattern is neither limited by atmospheric seeing nor the telescope diffraction limit, and its measurement
with large telescope
baselines provides angular resolutions that are
difficult to obtain from a single telescope.
In intensity interferometry the signal is
photon counts, a local measurement that
does not require the combination of light
from different telescopes,
which allows for
multiple and distant baselines that are technically difficult to achieve in Michelson interferometry.
Intensity interferometry also detects interference between spatially separated sources, making it inherently sensitive to the sky intensity distribution.

This study examines the telescope apertures and
baselines necessary for applying intensity interferometry to
measure  angular diameter distances to type~Ia supernovae
(SNe~Ia).
Such measurements will localize the three-dimensional position of SN host galaxies.
In addition,
distances to SNe~Ia, which serve as standardizable candles \cite{phillips1993absolute}, calibrate their absolute magnitudes, providing a direct pathway to measuring $H_0$.
We confine ourselves to the study of supernovae brighter
than $12$~mag, fainter than which would
stress the capabilities of foreseeable
next-generation observatories.
Our results scope the basic requirements for a purpose-built array designed for this science case.

\begin{figure}[htbp] %  figure placement: here, top, bottom, or page
   {\centering
   \includegraphics[width=3in]{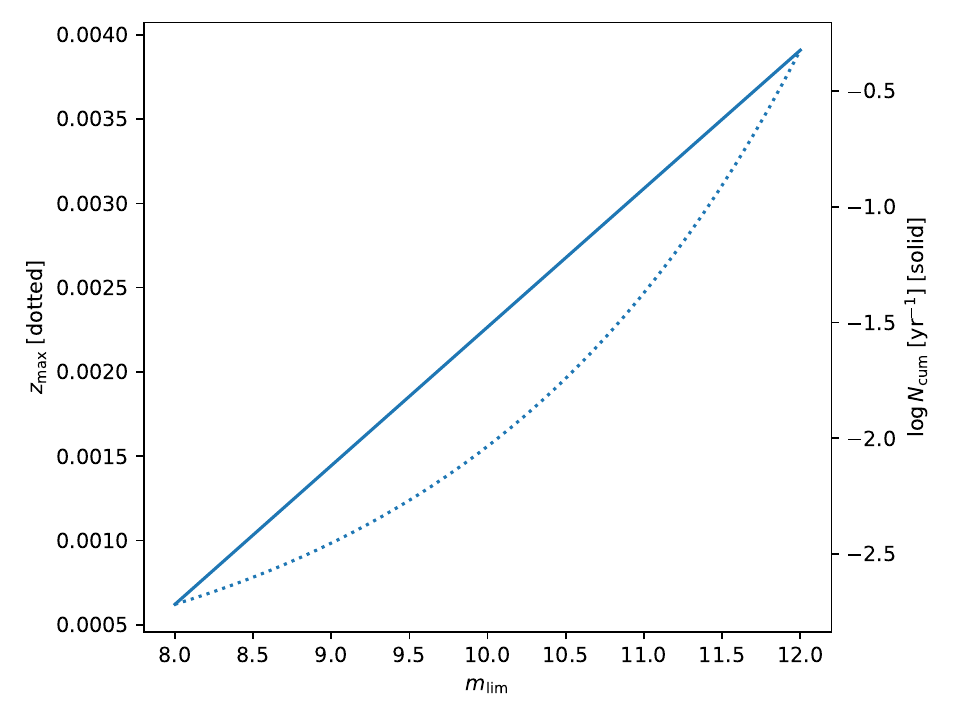} 
      \includegraphics[width=3in]{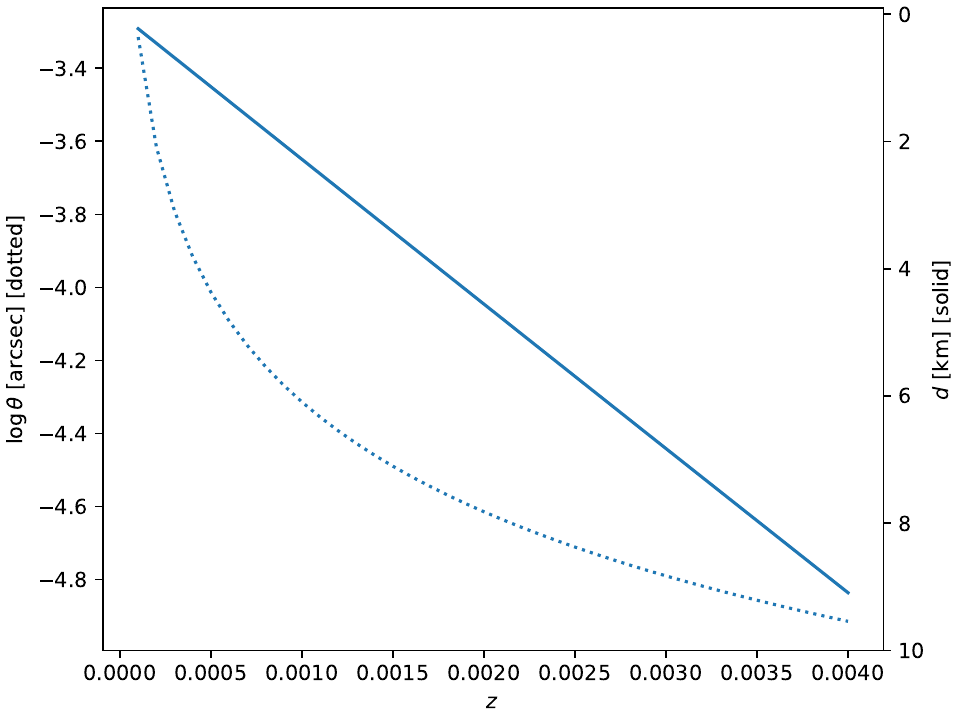}
      }
   \caption{Left:  Redshift depth (dotted line) and cumulative full-sky rate $N_\text{cum}$ (solid line) of SNe Ia 
   as a function of survey limiting magnitude $m_\text{lim}$.
   Right: Angular
   size at maximum light \added{(dotted line)} and telescope
   baseline \added{(solid line)} needed to resolve SNe~Ia
   at $\lambda=4400$\AA\  as a function of redshift.}

   \label{fig:rates}
\end{figure}

A reasonable number of  
$<12$~mag SNe are
available for the patient astronomer.
SNe~Ia have a rate of $r_v = 2.43
 \times 10^{-5}\,\text{SNe}\, \text{yr}^{-1} \text{Mpc}^{-3} h^3_
{70}$ \citep{2019MNRAS.486.2308F}
and on average reach a peak absolute magnitude
of $B =-19.13 - 5\log(h_{70})$ \cite{2002AJ....123..745R}.
These parameters
set the volume and thus the numbers of supernovae available to a survey with limiting magnitude
$m_\text{lim}$ assuming they are standard candles (see
Fig.~\ref{fig:rates}).
For $m_\text{lim}=12$~mag, the maximum detection depth is $z=0.004$ within which  $\sim 0.5$ SNe~Ia are expected annually over the full sky.
Since supernovae are not perfect standard candles, a differentially larger volume of intrinsically brighter but more distant objects can satisfy the detection limit.
A direct estimate of the SN~Ia rate based on a query to the Weizmann Interactive Supernova Data Repository \cite{2012PASP..124..668Y} gives 10 normal SNe~Ia with $m\le 12$~mag over the last 15 years, with increased frequency compared to prior years following the implementation of the PTF, iPTF, and ZTF 
\cite{2009PASP..121.1395L, 2018ATel11266....1K, bellm2019ztf}  transient
searches.
%  Given active searches covering both hemispheres, we expect to discover $\sim 1$ SNe~Ia each year with $m\le12$~mag at redshifts distributed around the curve in Fig.~\ref{fig:rates}.
% The location of the observatory dictates
% what fraction of these supernovae are visible.
Given active searches covering both hemispheres we expect the discovery of $\sim 1$ SNe~Ia each year, of which a significant subset will be accessible to a single intensity interferometry observatory.

Supernovae at maximum light at these distances can be spatially resolved with telescopes separated by $\lesssim 10$~km.
The angular
size of a supernova
at maximum is twice the
product of the 
the photospheric
velocity,
say taken from the blue trough of the signature
SiII P-Cygni profile in spectra,
and the time after explosion measured
from the light curve.
SNe Ia exhibit a photospheric velocity of
$\sim 10,000~\text{km}\,\text{s}^{-1}$
\cite{2005ApJ...623.1011B} and reach
peak brightness $\sim 20$ days after explosion \cite{2011A&A...528A.141J}.
The angular resolution
$\theta$ of optical intensity interferometers is set by the diffraction limit:
\begin{equation}
\theta
= 1.22 \frac{\lambda}
{\rho},
\label{eq:criterion}
\end{equation}
where $\lambda$ represents wavelength and $\rho$ the distance between telescopes.
Angular
sizes at maximum light and the required telescope baselines are plotted in
Fig.~\ref{fig:rates}
for $\lambda=4400$\AA.
Resolving SNe Ia out to
$z=0.004$  requires a baseline of less than
10~km.  The requisite distances are higher/lower before/after
peak brightness.

While core-collapse supernovae occur more frequently, their fainter intrinsic brightness confines them to a smaller volume. SNe Ia dominate magnitude-limited samples in the local universe, making them our focus.

This article explores the feasibility of measuring supernova angular sizes.  Section \ref{sec:profile} presents 
light-emission profiles of an SN~Ia at maximum light
generated by two  spectral synthesis codes.
The expected intensity interferometry signal due to these two
profiles are shown in Sec.~\ref{sec:signal}.  Section \ref{sec:angle}
provides the predicted precision in supernova distance for a fiducial telescope-instrument combination and 
includes a discussion on bias due to model inaccuracy and polarization.
Conclusions are given in Sec.~\ref{sec:conclusions}.

\section{Modeling the Supernova Emission Profile}
\label{sec:profile}
An understanding of the source's emission profile is needed to interpret intensity interferometry measurements, as they do not convey the phase information otherwise available when interfering electromagnetic fields.
Supernovae, as exploding stars, emit light through various physical processes occurring in an expanding medium. Their emission profile is wavelength-dependent and varies across different lines-of-sight toward the supernova surface, as atomic line transitions undergo Doppler shifts based on the velocity component directed toward the observer.
As such,
supernovae neither have a sharp edge nor 
have a constant emission over its surface.
When it comes to supernovae we do not think
of size, but rather 
a wavelength-dependent, nonuniform emission profile
whose angular extent is to be measured.

In this section we present the 
normalized emission profiles generated
by two spectral synthesis packages.
Both qualitatively reproduce observed
spectra, particularly the positions and shapes of the
P-Cygni profiles of several
atomic species.  
As such,
they offer an improvement over measuring
supernova size directly from the data.
Ongoing efforts seek to refine these codes' initial conditions and underlying physics, with the ultimate goal of achieving quantitative consistency between different models and observational data. 
We consider results from both packages, as they illuminate complementary aspects of the analysis.
Alternatively, simpler models
that focus on wavelengths corresponding to optically thick lines may suffice for our application.

The emission profiles provided by the models are denoted as $I_M(r_1,r_2)$, where $r=\sqrt{r_1^2+r^2_2}$ is 
the impact parameter in distance
units and the components $1$ and $2$ are for an arbitrary polar basis.
The intensity profile of the supernova
at distance $d$ in angular position on the sky  is
$I(\theta_1, \theta_2) \propto I_M(\theta_1 d,\theta_2 d)$.

\subsection{TARDIS}
A notable example in type Ia supernova modeling is the analysis 
of SN2011fe using TARDIS, a modular, open-source, steady-state, Monte Carlo radiative-transfer spectral synthesis code for models of supernova ejecta
\cite{2014MNRAS.440..387K}. TARDIS' strength is speed, which allows the fitting of supernova
parameters to generate an individualized model from data of a single supernova.
Speed comes at the expense of making several simplifying assumptions that do not fully apply to supernovae, including spherical symmetry,
no time dependence, and
a photosphere that sources the energy packets.
Despite these limitations, TARDIS has been used
to generate models that are consistent with spectroscopic
observations in the wavelength and 
phase regime relevant here.\footnote{A list of TARDIS-based publications is compiled at \url{https://tardis-sn.github.io/tardis/resources/research_done_using_TARDIS/research_papers.html}
} 
Recent applications demonstrate TARDIS's capabilities \cite{2020ApJS..250...12C, 2024ApJ...962..125C}, for example neural-network based modeling of supernova spectra has revealed correlations between radioactive $^{56}$Ni and
light curve properties of SNe and has 
derived the chemical abundances in the ejecta above the photosphere.

The example SN2011fe model's ejecta composition and density profile are the maximum likelihood sample from generative tomographic inference by O'Brien \textit{et al.} 2024 (in prep).  The ejecta density and nuclear decay products are evolved to the peak $B$-band time since explosion from \citet{2013A&A...554A..27P} with the corresponding input bolometric luminosity.  The inner photospheric velocity boundary is set such that the Rosseland mean optical depth is $\frac{2}{3}$ through the ejecta above.  TARDIS is run on the SN2011fe model with atomic data from Kurucz GFALL \citep{1995KurCD..23.....K} and line interactions are handled by the \texttt{macroatom} model \citep{2002A&A...384..725L}. Plasma ionization populations are solved through the \texttt{nebular} approximation \citep{1993A&A...279..447M} and approximate NLTE excitation populations are solved with the \texttt{dilute-lte} prescription \citep{1999A&A...345..211L}\footnote{See documentation for TARDIS at \url{https://tardis-sn.github.io/tardis/} for complete descriptions of configuration options}.
Line-intensity maps as a function of impact parameter are computed from the converged mean intensity in each line's extreme blue wing and the source function from the converged level populations and Einstein coefficients of the \textsc{TARDIS} simulation following the steps for the formal integral computation \citep{1999A&A...345..211L} but without performing the final integration step.  \replaced{Lines of sight}{ Intensity rays} 
are drawn though a spherically symmetric ejecta at each frequency and impact parameter, accumulating intensity with the optical depth within each shell, until exiting the ejecta.  A more thorough description of the procedure can be found in the TARDIS documentation
\cite{tardis_lucy}.
%\url{https://tardis-sn.github.io/tardis/physics/spectrum/formal_integral.html#calculating-i-nu-p}

\begin{figure}[htbp] %  figure placement: here, top, bottom, or page
   {\centering
   \includegraphics[width=3.5in]{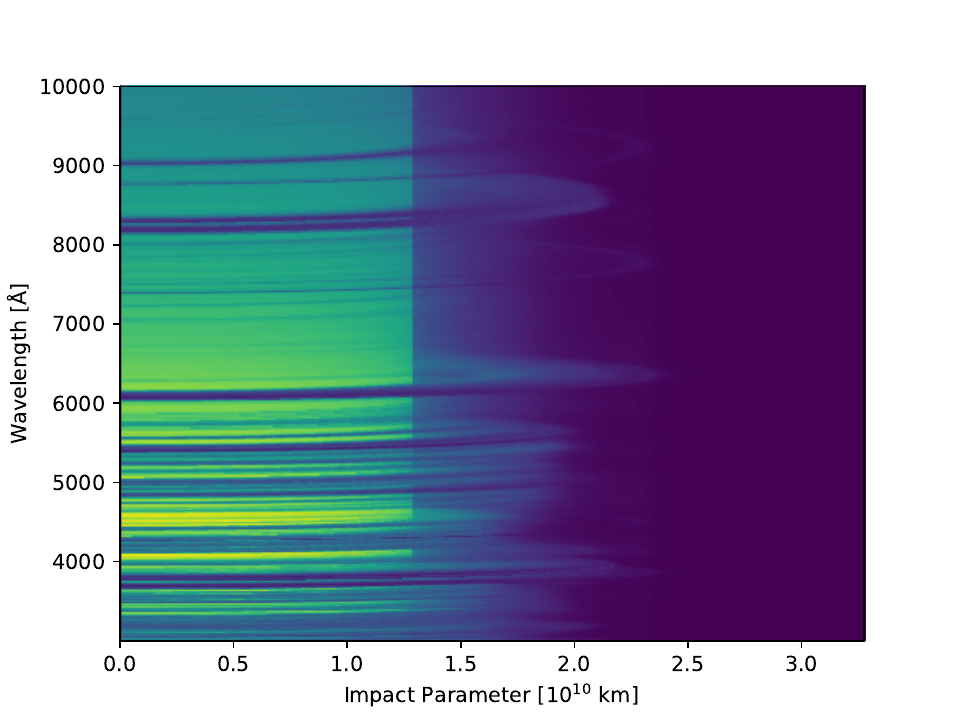}
   \includegraphics[width=3.5 in]{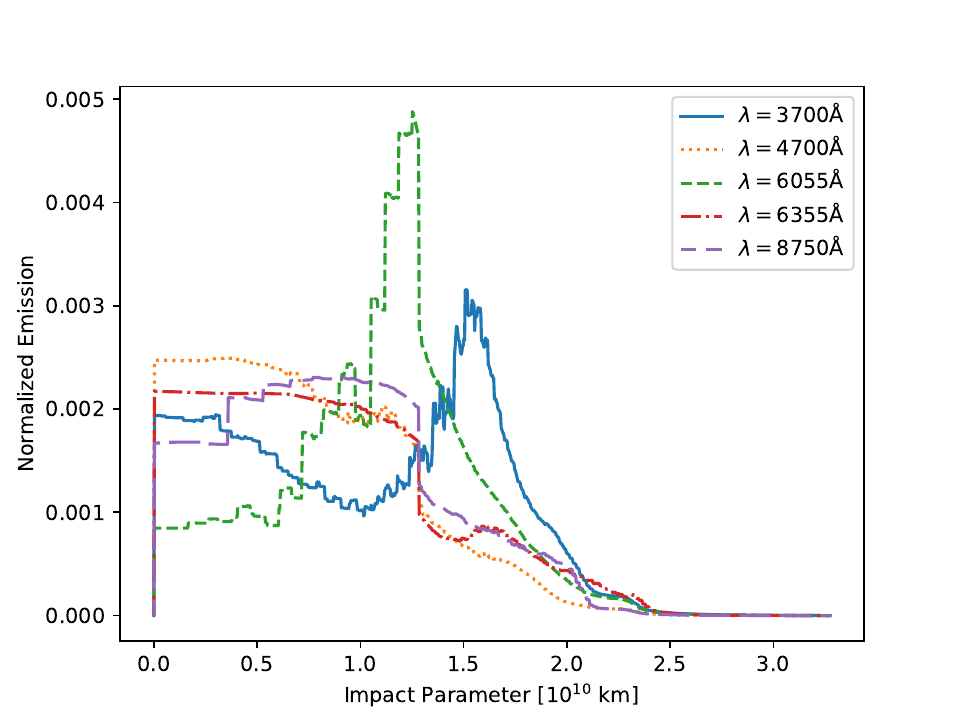} 
   }
   \caption{
   Left: Model emission of SN 2011fe at maximum light generated by TARDIS as a function of impact parameter along the line of sight. Right: Normalized emission as a function of impact parameter for the same model
    at select wavelengths that accentuate the contrast in radial profiles.}
   \label{fig:sn2011fe}
\end{figure}

\begin{figure}[htbp] %  figure placement: here, top, bottom, or page
   {\centering
   \includegraphics[width=4.5in]{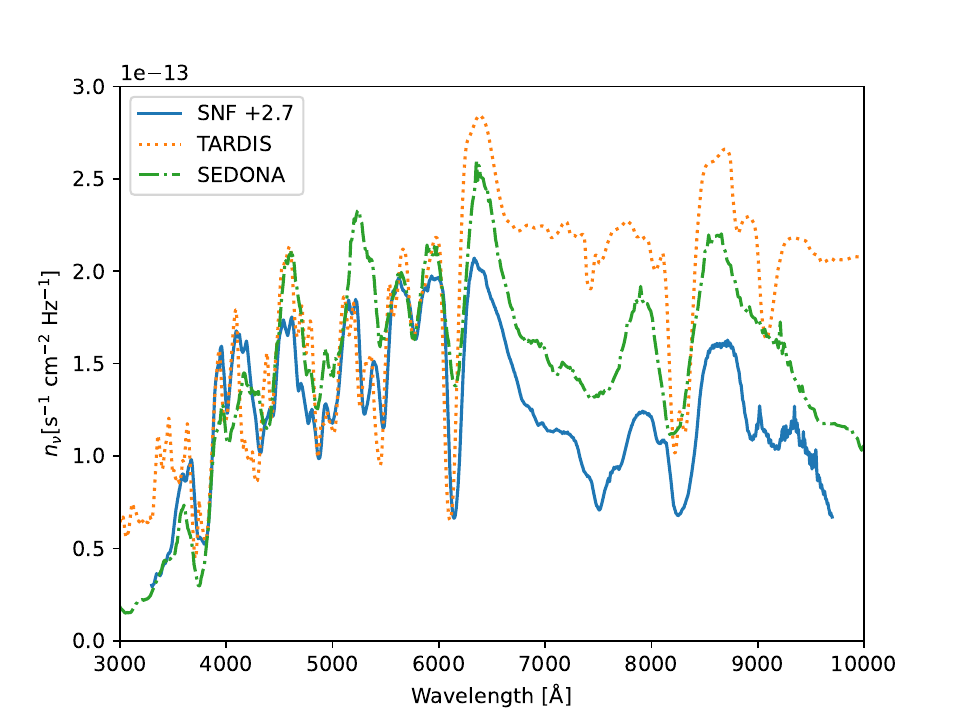} 
   }
   \caption{Type Ia supernova photon flux density per frequency at maximum light calculated by TARDIS and SEDONA models, normalized to $B$-band magnitude $B=12.0$. The observed spectrum from the Nearby Supernova Factory taken 2.7 days after maximum light is plotted for reference.}
   \label{fig:nlam}
\end{figure}

Figure~\ref{fig:sn2011fe}
displays the emission at maximum light for the SN2011fe model, showing emission as a function of 
impact parameter
$r$ (recall TARDIS assumes spherical symmetry).
In the same figure, selected wavelength slices illustrate the varying radial profiles characteristic of the expanding supernova atmosphere.
Figure~\ref{fig:nlam}
presents the emission spectrum integrated over the full supernova surface, which can be compared with observations by the Nearby Supernova Factory 2.7 days after maximum \cite{2013A&A...554A..27P}.
There is excess
flux calculated by TARDIS redward of 6500\AA, which will be
later shown to give excess 
intensity interferometry signal.

\subsection{SEDONA}

The SEDONA code solves the three-dimensional time-dependent radiative transfer problem in rapidly expanding supernova atmospheres including
$\gamma$-ray deposition from radioactive
materials to calculate
light curves, spectra, and polarization of aspherical supernova models 
 \cite{2006ApJ...651..366K}.  As such,
SEDONA provides a more precise model of supernovae
than TARDIS, though at greater computational expense.  A comprehensive comparison between TARDIS and SEDONA can be found in \citet{2022A&A...668A.163B}.

We use an upgraded version of SEDONA
to simulate the  N100 model, where
asymmetry is seeded by randomly placed ignition sites
\cite{2012ApJ...750L..19R}. 
The spectropolarimetry signal retrieval method is am integral based algorithm (IBT) \cite{Chen2024IBT}. 
The plasma excitation and ionization state is calculated using a spectral line data base with 35,344,426 spectral lines and 497,424 levels in total. 
The level population of Si, S, and Ca are calculated with nonlocal thermodynamic equilibrium (NLTE), other elements are calculated assuming local thermodynamic equilibrium (LTE). 
To accelerate the NLTE calculation in the 3D simulation, we utilize a deep-learning-based algorithm. A \replaced{convolutional}{ convolution} neural network (CNN), which predicts the absorption and emission coefficient using angle-averaged specific intensity ($J_{\nu}$) and elemental abundances as input, is trained and validated on the SEDONA radiative-transfer results of 119 1D SNe Ia models. CNN has reduced the NLTE computation time from 72 core seconds to 0.17 core seconds per zone, allowing us to perform a 3D time-dependent NLTE radiative transfer simulation. Details \replaced{will be}{ are} reported in Chen (in prep.).
 
\begin{figure}[htbp] %  figure placement: here, top, bottom, or page
   {\centering
   \includegraphics[trim={3cm 0 2cm 0},clip, width=7.5in]{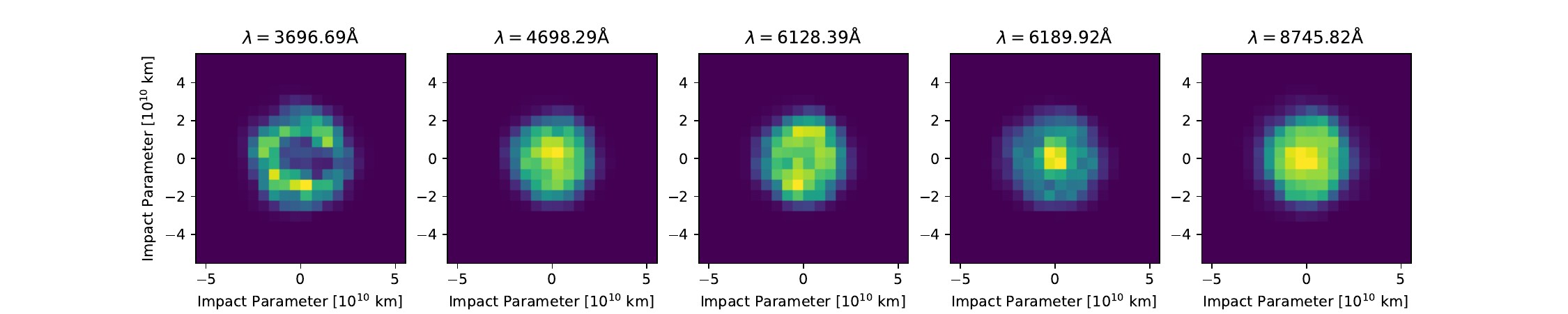}
   }
   \caption{SEDONA emission profiles 20 days after explosion (corresponding to the peak of the broadband light curve) at select wavelengths that accentuate the contrast in radial profles.}
   \label{fig:image}
\end{figure}

SEDONA generates time dependent and wavelength dependent emission profiles. 
Figure~\ref{fig:image}
displays profiles at maximum light for selected wavelengths chosen to illustrate diverse emission behaviors.  These wavelengths differ from those used in the TARDIS simulation to highlight SEDONA-specific features.
Like the TARDIS simulation, the profiles show clear radial dependence.
However, SEDONA reveals an additional characteristic: the intensities lack axial symmetry.
Figure~\ref{fig:nlam}  presents the spectrum at maximum light integrated over the complete supernova surface.

\section{Normalized Visibility}
\label{sec:signal}
The supernova's profile in angular position on the sky is denoted as
$I(\lambda, \theta_1, \theta_2)$.  The 
intensity interferometry signal is proportional to
$\mathcal{V}^2$, where
the normalized visibility is defined
as the Fourier transform
\begin{equation}
    \mathcal{V}(\lambda, u,v) \overset{\mathscr{F}}\Leftrightarrow \frac{I(\lambda, \theta_1,\theta_2)}{\int I(\lambda, \theta_1,\theta_2) d\theta_1 d\theta_2},
\end{equation}
where the
Fourier transform variables $u$--$v$ 
span a coordinate system
that gives the relative telescope positions
on a plane perpendicular to the line of sight.
For example, an intensity profile $I$ that is a uniform circle with angular
diameter $\theta$
yields the Airy profile
\begin{equation}
\mathcal{V}^2(u,v) = 
\left(\frac{2J_1(\zeta)}{\zeta}\right)^2,
\label{eq:airy}
\end{equation}
where $\zeta = \pi \rho \theta/\lambda$ and $\rho=\sqrt{u^2+v^2}$;
this relationship gives rise to the resolution condition
 $\theta
= 1.22 \lambda/
{\rho}$.

\begin{figure}[htbp] %  figure placement: here, top, bottom, or page
   {\centering
\includegraphics[width=4.5in]{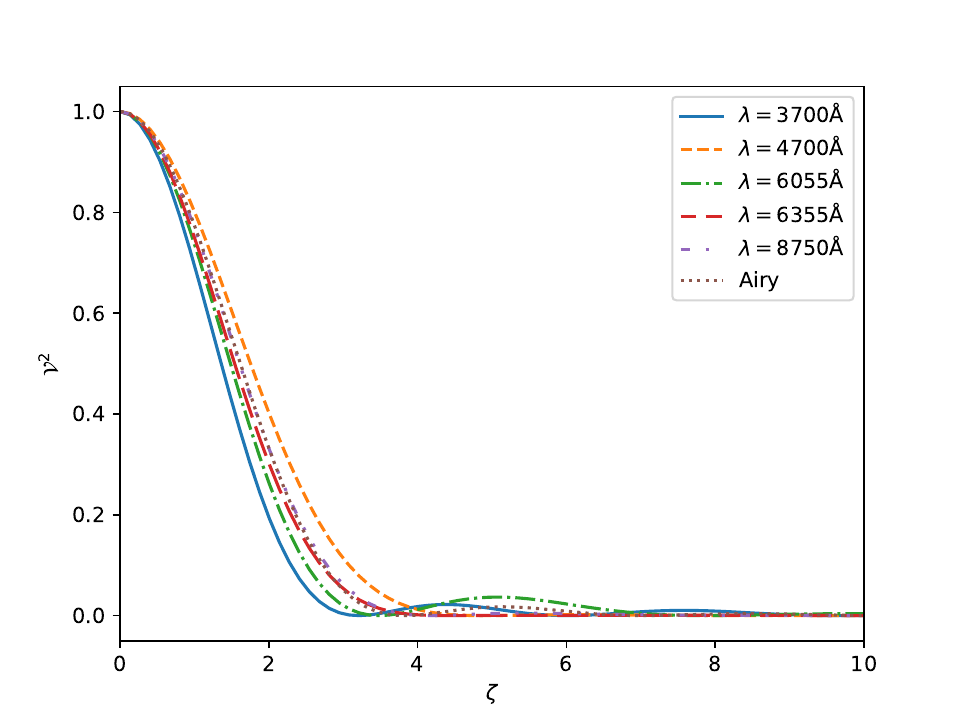} }
   \caption{
   $\mathcal{V}^2$ for the TARDIS emission profile as a function of telescope separation at various wavelengths, where $\zeta = \pi \rho \theta/\lambda$ and $\theta$ corresponds to $2$ times the impact parameter of $0.875 \times 10^{10}$ km. The Airy disk profile is shown for reference.
   }
   \label{fig:slices}
\end{figure}

As shown in Sec.~\ref{sec:profile}, supernova expansion creates a nonuniform emission profile, resulting
in $\mathcal{V}^2$ that varies with wavelength and deviates from an Airy disk. 
In the TARDIS model, the axially symmetric emission profile produces a corresponding symmetric visibility that depends on telescope separation.
Figure~\ref{fig:slices} presents $\mathcal{V}^2$
for select wavelengths.
While the expected signal differs from an Airy disk— demonstrating the value of spectral synthesis codes for improved precision — it resembles one sufficiently to justify applying the resolution criterion from Eq.(\ref{eq:criterion})
to supernovae.

\begin{figure}[htbp] %  figure placement: here, top, bottom, or page
   {\centering
   \includegraphics[trim={3cm 0 0 0},clip,width=7in]{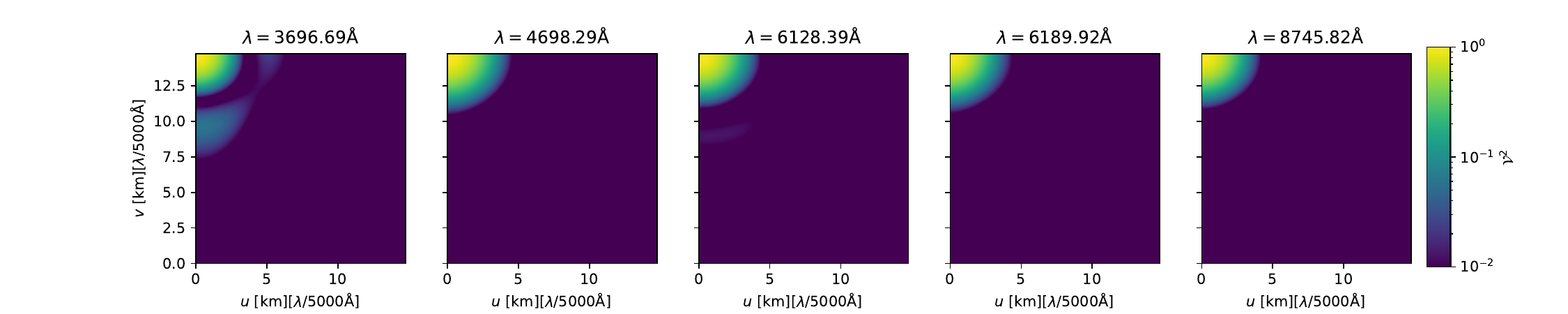} 
   }
   \caption{
   SEDONA model maps of the intensity interference signal $\mathcal{V}^2$
as function of relative telescope position $u$--$v$ at select wavelengths for a supernova at z=0.004.}
   \label{fig:sedona_map}
\end{figure}

\begin{figure}[htbp] %  figure placement: here, top, bottom, or page
   {\centering
   \includegraphics[width=7.5in]{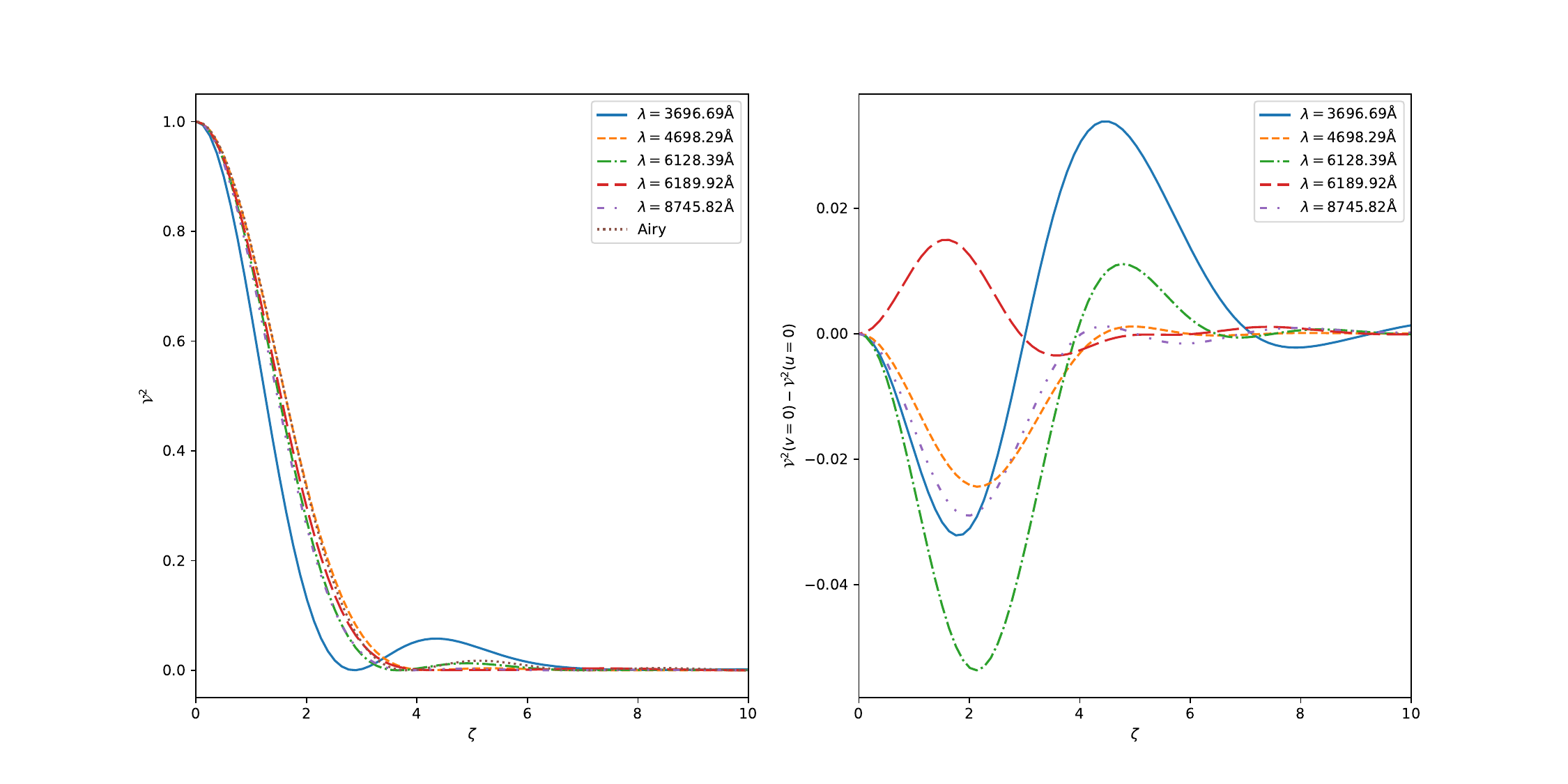}
   }
   \caption{ $\mathcal{V}^2$ for the SEDONA profiles at selected wavelengths.  Left:
   Slices for $v=0$ with
   the Airy profile is shown for reference.
   Right: Difference between  two slices $\mathcal{V}^2(v=0)-\mathcal{V}^2(u=0)$.}
   \label{fig:gamma2_sedona}
\end{figure}

For the SEDONA emission maps, $\mathcal{V}^2$ calculations reveal wavelength dependent patterns. 
Figure~\ref{fig:sedona_map} displays $\mathcal{V}^2$ for select wavelengths,
while Figure~\ref{fig:gamma2_sedona} shows
$\mathcal{V}^2$ slices at $v=0$.
As with TARDIS, these profiles exhibit wavelength dependent deviations from an Airy profile. The asymmetric profile manifests in differences at the level of $\lesssim 0.05$.

\section{Distance Precision}
\label{sec:angle}
In intensity interferometry a single telescope measures a series of photon counts $N$ in a narrow frequency bin.
 The key measurement statistic is
the  covariance
between 
the time-dependent counts from two telescopes $\langle\delta N_1 \delta N_2 \rangle$.
For an unpolarized source with specific flux 
$F_\nu$ at central frequency $\nu_0$ and bandwidth $\Delta \nu$, and
given a total observing time $T_{\text{obs}}$,
the signal-to-noise is
\begin{equation}
    \text{SNR} = \frac{\mathcal{V}^2}{\sigma_{
    \mathcal{V}^2}}
    \label{eq:snr}
\end{equation}
in the Poisson-noise limit
and when the detector timing jitter $\sigma_t$ satisfies
$\sigma_t \Delta \nu \gg 1$, and where
\begin{equation}
\sigma_{\mathcal{V}^2}^{-1} =
    \frac{d\Gamma}{d\nu}\left(\frac{T_\text{obs}}{\sigma_t}\right)^{1/2} (128\pi)^{-1/4}
\end{equation}
\cite{2024PhRvD.109l3029D}.
Here $d\Gamma/d\nu=\epsilon AF_\nu/(h\nu_0) = \epsilon A n_\nu$
(proportional to the
spectra plotted in Fig.~\ref{fig:nlam}), 
$\epsilon$ is the total system throughput, and
$A$ is the effective telescope area.
To simplify notation, we consider
the ``signal'' to be
$\mathcal{V}^2$ rather than
$\langle\delta N_1 \delta N_2 \rangle$.

The noise has the feature
of being independent of  bandwidth
when dominated by timing jitter,
when the detector cannot resolve the higher-frequency temporal
coherence associated with broader bandpasses.
For a detector jitter of 30 ps FWHM,
this condition is 
$R \ll 7000$ at optical wavelengths.
We can thus consider light
filtered at
a  resolution  ($R> 100$) that does not appreciably
smear the interference pattern of the central
frequency.
Note that this feature 
 allows us to disperse the light without
noise degradation, enabling  multiplex advantage through simultaneous
timing measurements at different wavelengths
with a detector array.

The emission profile projected onto the sky is given by
$I(\theta_1, \theta_2) \propto I_M(\theta_1 d,\theta_2 d)$, where $d$ represents the supernova distance.  For computational convenience, we parameterize the distance as $d = s d_0$ and
where $d_0$ is a fiducial distance and
identify
$I_0(\theta_1, \theta_2) \propto I_M(\theta_1 d_0,\theta_2 d_0)$ and its corresponding visibility $\mathcal{V}_{I_0}$.  
The parameter $s$ is the relative distance with respect to $d_0$.
The model for the emission profile  can then
be expressed as
$I(\theta; s) \propto I_0(s\theta)$ with its Fourier transform's square representing the model signal.

With the model signal and measurement uncertainty established, we can calculate the uncertainty in relative distance to the supernova using the Fisher matrix formalism. We analyze the TARDIS and SEDONA profiles separately to highlight the distinctions between axially symmetric and asymmetric profiles.

\subsection{TARDIS}
\label{sec:snrtardis}

The intensity profile for the supernova
at distance $d=s d_0$ 
is given by $I(\theta; s) \propto I_0(s\theta)$.
The normalized visibility is
\begin{align}
    \mathcal{V}_I(\rho;s)
     & =  \mathcal{V}_{I_0}\left(\frac{\rho}{s} \right),
\end{align}
where $\rho = \sqrt{u^2+v^2}$
and axial symmetry is assumed.

The derivative of the expected signal
$\mathcal{V}_I^2$ with respect to $s$
can be expressed in terms of the fiducial profile
as follows
\begin{align}
    \frac{d \mathcal{V}_{I_0}}{d \rho} (\rho) & = - 
    (2\pi)^2 \int_0^\infty I_0(\theta) J_1(2\pi\rho\theta)\theta^2 d\theta \\
    \frac{d \mathcal{V}_I}{d s} (\rho)&=
  -
    \frac{1}{s} \rho' \frac{d \mathcal{V}_{I_0}}{d \rho'}(\rho' )\\
    \frac{d \mathcal{V}^2_I(\rho)}{d s} &=
    2 \mathcal{V}_I^* \frac{d \mathcal{V}_I(\rho)}{d s},    \end{align}
where $\rho'=\rho/s$.
Using the above, the Fisher
estimate for the SNR of $s$  is
\begin{align}
    \text{SNR}_s 
    & = \sigma_{\mathcal{V}^2}^{-1} \left| 2 \mathcal{V}^*_I \frac{d \mathcal{V}_I}{d s}\right| .
\end{align}

\begin{figure}[htbp] %  figure placement: here, top, bottom, or page
   {\centering
   \includegraphics[width=4.5in]{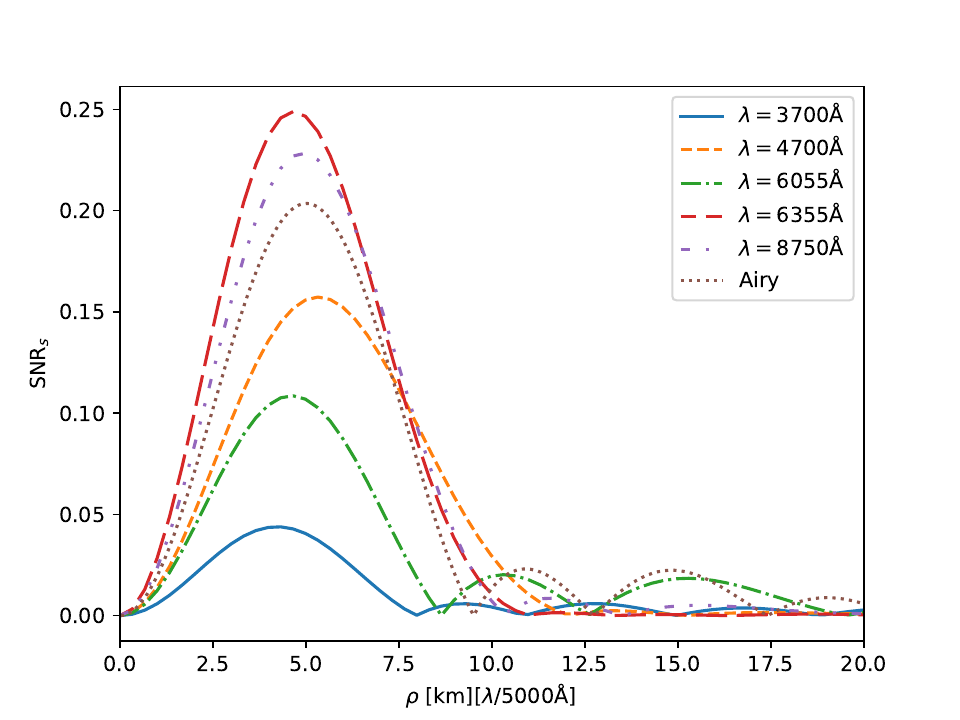}
   }
   \caption{Signal-to-noise of  distance parameter $s$ as a function of telescope
   separation $\rho$
    at selected wavelengths for the TARDIS model.  
     Measurements assume a supernova at $z=0.004$ with $B=12$ mag, using two 
    Keck-like telescopes with effective
     throughput
     $\epsilon=0.39$ and a 1 hour exposure time. The signal-to-noise function for a circular profile with radius $0.875 \times 10^{10}$ km is shown for reference.}
   \label{fig:snr}
\end{figure}

For the case of a circular aperture where
the signal for a telescope pair is an Airy disk of Eq.(\ref{eq:airy}),
\begin{align}
\mathcal{V} & =\frac{2J_1(\zeta)}{\zeta} \\
    \frac{d\mathcal{V}_s}{ds} & = -\frac{1}{s} \zeta'   \left( \frac{J_0(\zeta')}{\zeta'}- \frac{J_2(\zeta')}{\zeta'} -\frac{2J_1(\zeta')}{\zeta'^2}\right)
\end{align}
so
\begin{align}
    \text{SNR}_s 
    & = {2} \sigma_{\mathcal{V}^2}^{-1} \left|  \frac{2 J_1(\zeta')}{\zeta'} \left( {J_0(\zeta')}- {J_2(\zeta')} -\frac{2J_1(\zeta')}{\zeta'}\right) \right|,
\end{align}
where $\zeta'=\zeta/s$.  The shape of this function
is plotted in Fig.~\ref{fig:snr}.

We adopt two Keck-like telescopes
separated by a baseline distance $\rho$ as a reference for our signal-to-noise calculations.
Keck has an effective aperture of a 9.96~m diameter telescope with mirror smoothness at
a level of a few nm.  We assign a combined
mirror reflectivity of 0.73, filter transmission of 0.9, and a detector quantum efficiency of 0.6
to give an effective
total throughput of $\epsilon = 0.39$.
Each telescope is equipped with photon counters with a timing jitter of 30 ps FWHM or $\sigma_t=13$~ps RMS.  For the moment only
a single wavelength channel is considered,
although spectral multiplexing is  possible.  

Our reference Keck-like telescope pair can be translated
to a larger number of small-aperture telescopes.
The signal-to-noise from single baseline is proportional to
the collecting area.  The combined signal-to-noise of
$B$ baselines is proportional
to $\sqrt{B}$, where the number of baselines
is $B={N \choose 2}$ and $N$ is 
the number of telescopes.
With these scalings, the signal-to-noise
of two 10m telescopes is achieved with
566 0.5m, 
142 1m, or 36 2m telescopes.
Alternatively, distributing the primary mirror area from two large aperture telescopes to
increasing numbers of smaller telescopes
improves the signal-to-noise ratio by up to a factor $\sqrt{2}$.

The source is modeled as a supernova at $z=0.004$
with
$B=12$~mag and an emission profile
given by TARDIS.
These parameters, together with the exposure time
and detector jitter, determine the noise
$\sigma_{\mathcal{V}^2}$.
The signal to noise ratio of the $s$ parameter
as a function of telescope separation
computed for a one hour observation
$T_\text{obs} = 1$~h exposure is shown in Fig.~\ref{fig:snr}.
The distance is best constrained with telescope separations where the gradient
in the signal is steepest and
at wavelengths where $n_\nu$ is highest,
with the caveat that
TARDIS overestimates flux for wavelengths redward of 6500~\AA\ as shown in Fig.~\ref{fig:nlam}.
The signal to noise can be increased with multiple exposures and wavelength multiplexing.
For signal-to-noise ratios of order 0.15 per measurement, a factor
of $133^2$ independent
measurements are needed to achieve a total $\text{SNR} \approx 20$ enabling 
a 5\% precision on the distance,
comparable to the
intrinsic dispersion from SNe~Ia.

\subsection{SEDONA}
In this section we consider the
SEDONA calculated emission profile, which lacks
axial symmetry.
We will show that the statistical
size constraints approach those
of the one-dimensional model considered in the previous
section.
The more important point of focus are the features that arise
when the supernova emission is not symmetric.

The model intensity profile requires an extra 
 parameter $\phi$ to specify its orientation in the sky.
We again consider distances relative
to an initial
estimate $d_0$, so that
distance parameter is $s=d/d_0$. 
The intensity profile for the supernova
is expressed
as $I(\theta_1, \theta_2; s, \phi) \propto I_0\left(s(\theta_1\cos{\phi}-\theta_2\sin{\phi}),s(\theta_1\sin\phi+ \theta_2\cos{\phi}\right))$ and
\begin{align}
    \mathcal{V}_I(u,v;s)
     & =  \mathcal{V}_{I_0}\left(\frac{u\cos{\phi} - v \sin{\phi}}{s},\frac{u\sin{\phi} + v\cos{\phi}}{s} \right).
\end{align}

The partial derivatives of 
$\mathcal{V}_I$ with respect to $s$
and $\phi$
can be expressed in terms of the fiducial profile
as follows
\begin{align}
    \frac{\partial \mathcal{V}_{I_0}}{\partial u} (u,v) & = -i 2\pi
    \int \int \theta_1 I_0(\theta_1,\theta_2) e^{-i2\pi (u\theta_1 + v\theta_2)}d\theta_1 d\theta_2\\
    \frac{\partial \mathcal{V}_I}{\partial s} (u,v)&=
  -
    \frac{1}{s} \left( u' \frac{\partial \mathcal{V}_{I_0}}{\partial u'}( u',v') +  v' \frac{\partial \mathcal{V}_{I_0}}{\partial v'}( u',v') \right)\\
    \frac{\partial \mathcal{V}_I}{\partial \phi} (u,v)&=
  -v' \frac{\partial \mathcal{V}_{I_0}}{\partial u'}( u',v') + u' \frac{\partial \mathcal{V}_{I_0}}{\partial v'}( u',v') ,    \end{align}
where $u'=(u\cos\phi - v\sin\phi)/s$ and $v'=(u\sin\phi + v\cos\phi)/s$.
The Fisher matrix is
\begin{equation}
    F = \sigma_{\mathcal{V}^2}^{-1} 
    \begin{pmatrix}
     \sum \left(\frac{\partial \mathcal{V}_I^2}{\partial s}\right)^2 &  \sum\frac{\partial \mathcal{V}_I^2}{\partial s}  \frac{\partial \mathcal{V}_I^2}{\partial \phi}\\
     \sum\frac{\partial \mathcal{V}_I^2}{\partial s}  \frac{\partial \mathcal{V}_I^2}{\partial \phi} & \sum \left(\frac{\partial \mathcal{V}_I^2}{\partial \phi}\right)^2
    \end{pmatrix},
\end{equation}
where the summation is over independent measurements. The
estimate for the SNR of $s$ is
\begin{align}
    \text{SNR}_s 
    & = \left( F^{-1}_{ss} \right)^{-1/2} .
\end{align}

At least two pairs of measurements are necessary to measure the two parameters of
the model.  Two observing strategies
are considered. In the first strategy 
there are two baselines perpendicular to each other.  Each baseline has a
thirty minute exposure, resulting in a total exposure time of one hour for both pairs.
The second  strategy uses 
three baselines  obtained from three telescopes positioned at the vertices of a right isosceles triangle.  Two
pairs have the same separation and orientations 90$^\circ$ relative to each other, while the third pair has a
45$^\circ$ orientation and a $\sqrt{2}$ separation relative to the others.
In this case, each pair has a
twenty minute exposure again giving a total exposure time of one hour.
Both strategies use the same telescopes and detectors described in  Sec.~\ref{sec:snrtardis}.  

\begin{figure}[htbp] %  figure placement: here, top, bottom, or page
  { \centering
   \includegraphics[trim={3cm 1cm 0 0},clip,width=7in]{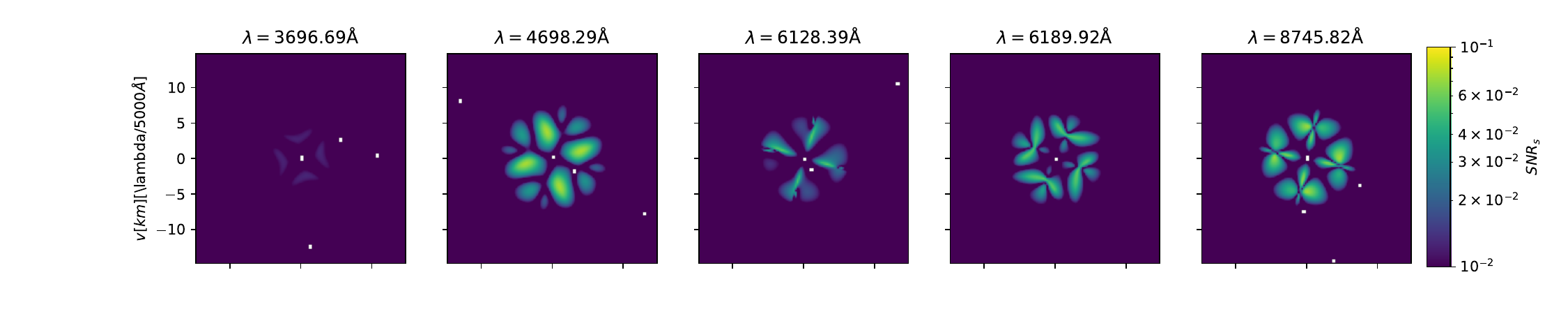} 
   \includegraphics[trim={3cm 1cm 0 1cm},clip,width=7in]{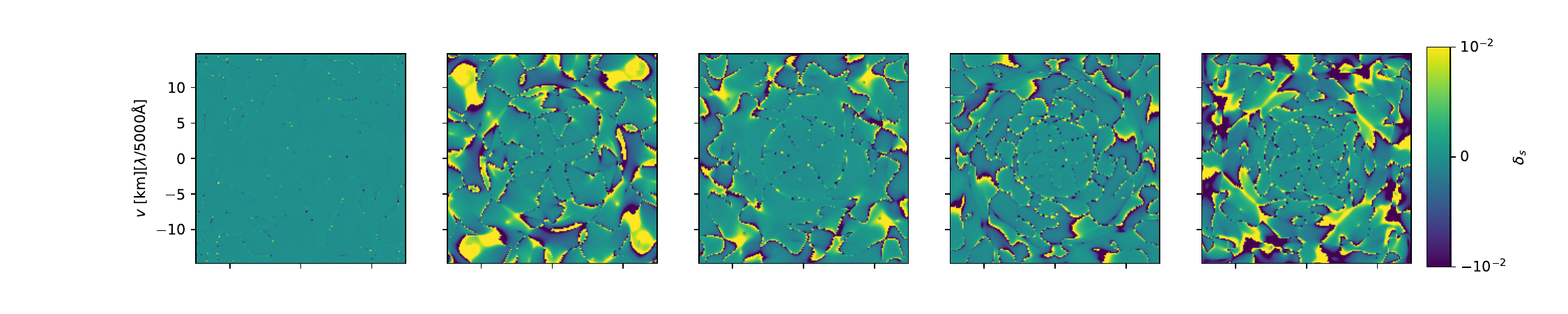} 
   \includegraphics[trim={3cm 1cm 0 1cm},clip,width=7in]{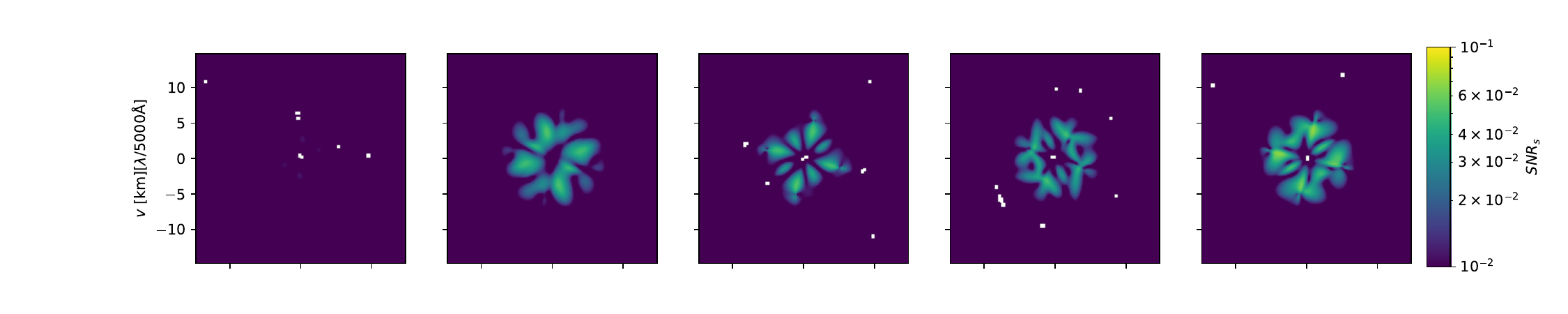} 
   \includegraphics[trim={3cm 0 0 1cm},clip,width=7in]{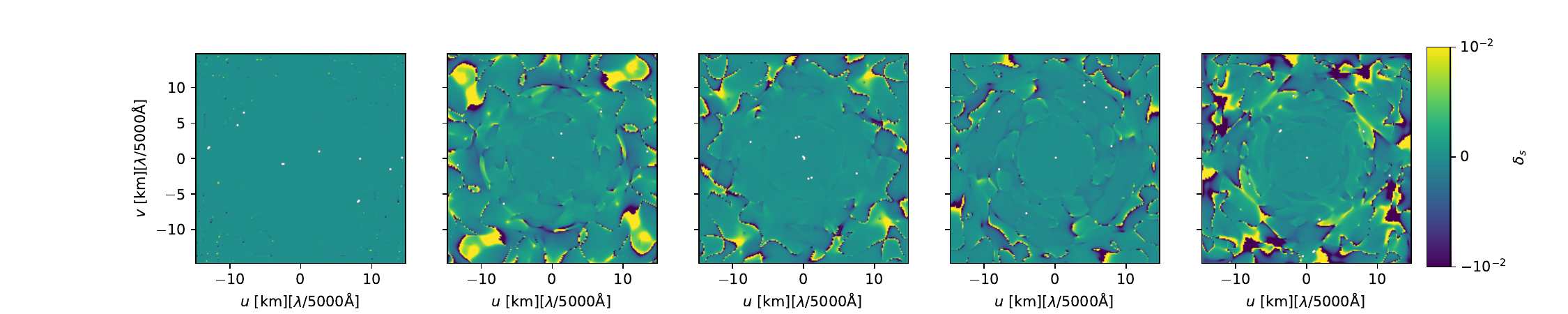} 
   }
   \caption{
   Results for distance parameter $s$ measurements
   using SEDONA model with the two-pair configuration (top two panels) and the three-pair configuration (bottom two panels) at select wavelengths.
    Assumes a supernova at $z=0.004$
    with $B=12$~mag.
    First and third panels: Signal-to-noise ratio of the relative distance parameter $s$.
    Second and fourth panels: Biases in relative distance $\delta_s$.
    The $u$-$v$ coordinate corresponds to one
of the baselines with equal separation.}
   \label{fig:sedona}
\end{figure}

The resulting signal-to-noise maps for selected wavelengths are presented in Fig.~\ref{fig:sedona}.
The $u$-$v$ coordinate corresponds to one
of the baselines with equal separation.
White points in the maps indicate locations where the Fisher matrix inversion encounters numerical exceptions, which occurs when derivatives
with respect to $\phi$ are small. (An axially symmetric
profile such as from TARDIS would lead to singular Fisher matrices.)

The effect of intensity uncertainty is explored by randomly perturbing the emission of each spatial element by 5\%, which
propagates into a signal bias $\Delta \mathcal{V}^2$.  The same perturbation realization is applied to all wavelengths.
The relative distance bias $\delta_s$ caused by $\Delta \mathcal{V}^2$ is
given by the Fisher formalism \cite{2006APh....26..102L} as:
\begin{equation}
\delta_s = F^{-1}_{ss} \sum_k
\frac{\partial \mathcal{V}^2_k}{\partial s} \sigma_{\mathcal{V}}^{-2} \Delta \mathcal{V}^2_k
+ F^{-1}_{s\phi}\sum_k
\frac{\partial \mathcal{V}^2_k}{\partial \phi} \sigma_{\mathcal{V}}^{-2} \Delta \mathcal{V}^2_k.
\end{equation}
The biases for both configurations are calculated and
shown in Fig.~\ref{fig:sedona}.
The largest biases occur in regions with low signal-to-noise ratios.

The axial asymmetry of the supernova profile results in measurement precisions and biases that depend nontrivially on telescope pair orientations.  The signal-to-noise map for the three-pair configuration is more evenly distributed (blurred) compared to the two-pair configuration. Additionally, the bias map for the three-pair configuration is less speckled and has lower amplitude. Increasing the number of correlation pairs with independent orientations helps mitigate systematic effects caused by profile asymmetry.

The fluctuations that give rise to the cross-correlation signal arise from fluctuations in the number of photons in each state that
make up the incident wave. 
The derivation of Eq.(\ref{eq:snr}) assumes
unpolarized light, where the two polarization
states with the same frequency have equal intensity.
For partially polarized light, the count
correlation of each polarization state
must be calculated separately, and
then combined to give the total signal.

\begin{figure}[htbp] %  figure placement: here, top, bottom, or page
   {\centering
   \includegraphics[trim={3cm 0 0 0},clip, width=7.in]{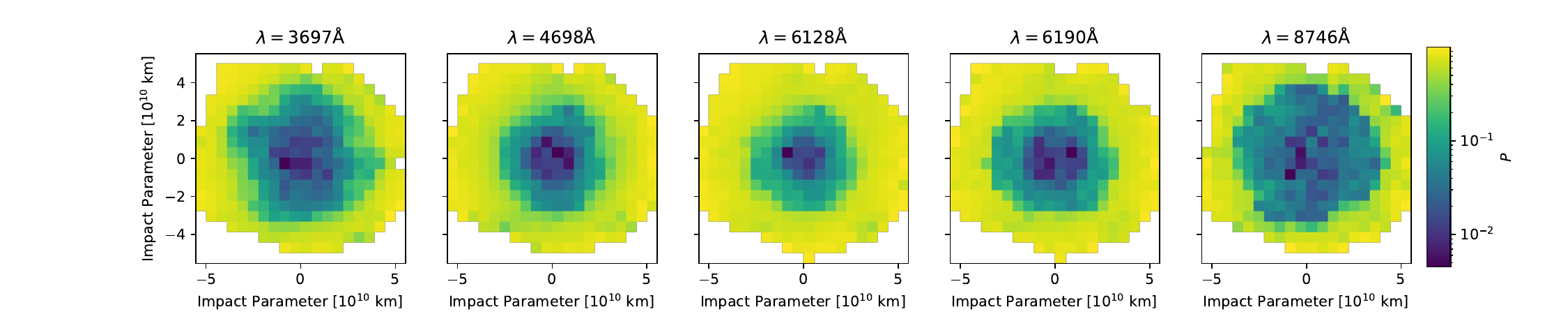}
   \includegraphics[trim={3cm 0 0 1cm},clip, width=7.in]{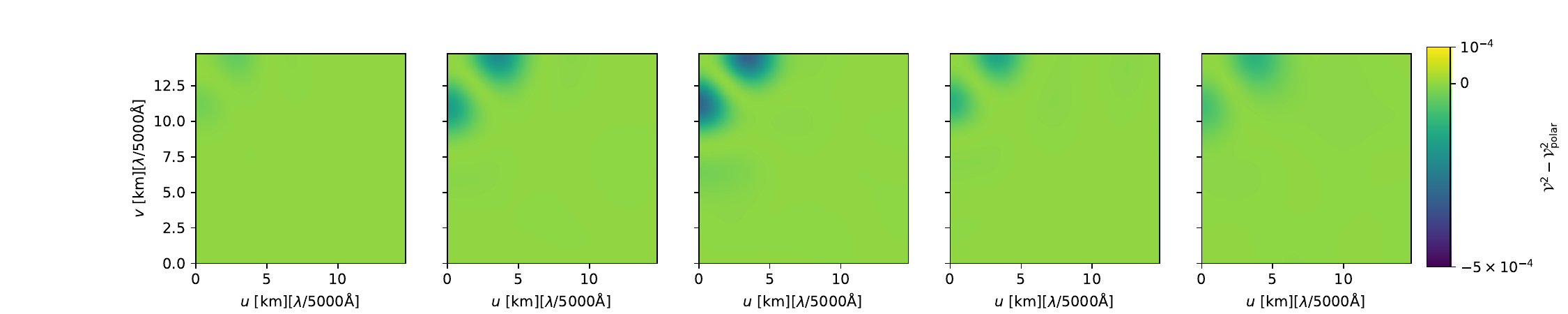}}
   \includegraphics[trim={3cm 0 0 1cm},clip,width=7in]{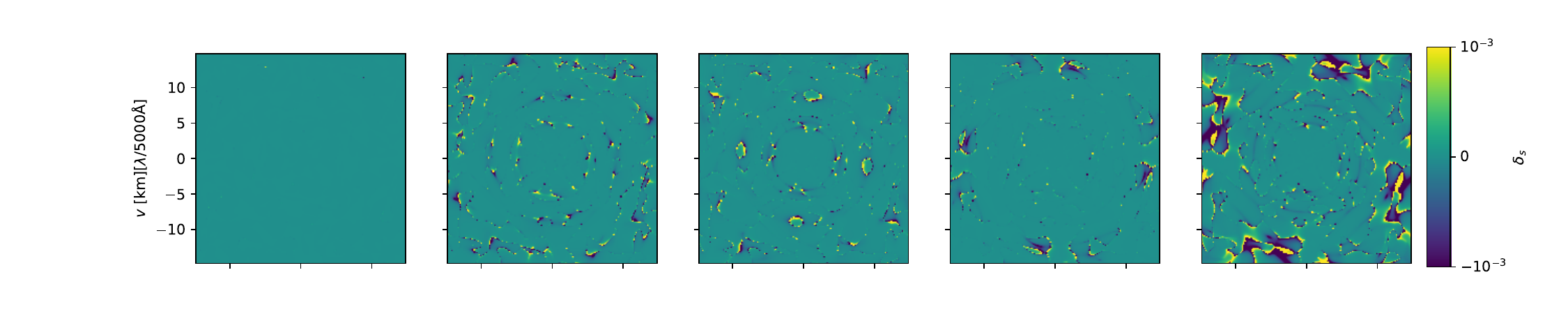} 
   \includegraphics[trim={3cm 0 0 1cm},clip,width=7in]{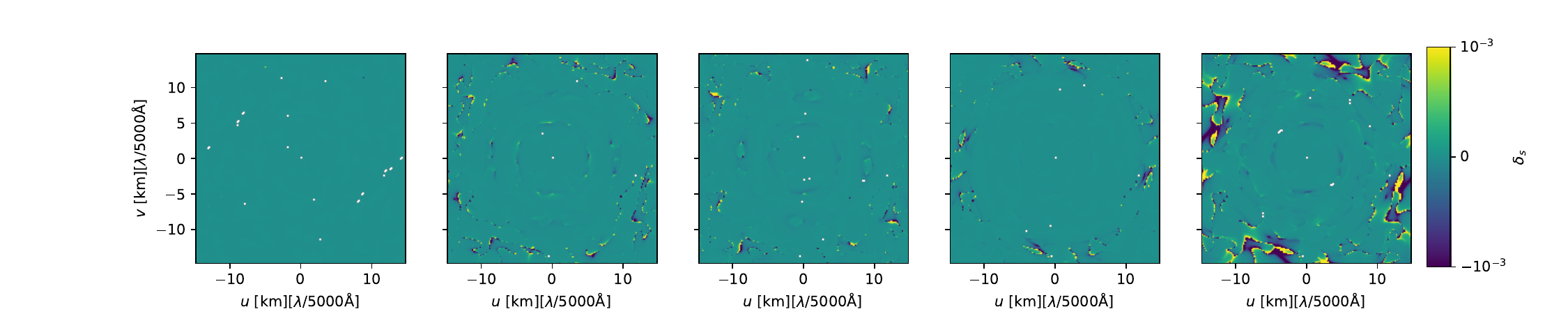} 
   \caption{Top panel: SEDONA polarization maps 20 days after explosion (at broadband light curve peak) at select wavelengths for a supernova at $z=0.004$. Second panel: Difference between expected signals with and without accounting for polarization, $\mathcal{V}^2 - \mathcal{V}^2_\text{polar}$. 
Difference between expected signals with and without accounting for polarization, $\delta_s$,
for the two-pair (third panel) and three-pair (fourth panel) configurations, respectively.
The $u$-$v$ coordinate corresponds to one
of the baselines with equal separation.}
   \label{fig:P_sedona}
\end{figure}

The integrated
polarization of SNe~Ia is observed to be $P\lesssim 0.005$, except at wavelengths corresponding to high-velocity Ca II, where it can be higher \cite{2008ARA&A..46..433W}.
SEDONA calculates similar integrated polarization with a nontrivial spatial structure. Polarization maps generated by SEDONA 
(Fig.~\ref{fig:P_sedona})
show weak polarization toward the core, where the supernova is brightest, and strong polarization at the outer fringes, where emission is weakest. 
The  difference between the intensity interferometry signals
calculated with and without considering polarization,
$\mathcal{V}^2 - \mathcal{V}^2_\text{polar}$,
is also shown
in Fig.~\ref{fig:P_sedona}.
Extreme differences are localized in $u$-$v$ space
and are  $<5\times10^{-4}$.
The relative distance bias $\delta_s$ for both two-
and three-pair configurations is $ \lesssim 0.005$ in regions with high signal-to-noise ratio.

\section{Conclusions}
\label{sec:conclusions}

We have demonstrated the feasibility of using intensity interferometry at optical wavelengths to achieve high-precision measurements of supernova distances, leveraging upcoming observatories and advanced detectors. This approach relies on telescope arrays with large light-gathering areas but moderate optical quality, and fast, multiplexed photon detectors. Given the expected supernova rate, multiple events can be observed within the survey volume during the lifetime of an observing campaign.

For individual SNe Ia, distance measurements combined with multiband light curves enable the determination of the Hubble constant,
$H_0$.
 The
supernova distance $d$ (measured in Mpc) and observed magnitude $m$
are used to calculate the absolute magnitude
through the relation
$M = m + 5\log{d} + 25$.  
Independently, light curve and/or spectroscopic data together with the SN~Ia Hubble diagram are used to determine the SN's value of $\mathcal{M}$, using tools like the SALT2 light-curve model
and magnitude calibration \cite{2007A&A...466...11G}. The Hubble
constant is then derived using the relation
$\mathcal{M} = M - 5\log{H_0}$.

Further work is needed to quantify model-specific systematic errors. Comparing TARDIS and SEDONA predictions for the same supernova across common phases and viewing angles serves as a robustness check. Identifying wavelengths that are less sensitive to modeling choices would be particularly beneficial.

Much of the current activity \added{in} intensity
interferometry is motivated by the availability of
current (VERITAS, MAGIC, H.E.S.S.) and future Cherenkov telescopes, \replaced{i.e.\ the Cherenkov Telescope Array Observatory}{ (CTA South MST)}.
Unfortunately, these telescopes are not capable of supporting
the assumptions made in this article.
Their coarse $>$arcmin point-spread function results in
contamination from the background galaxy and other projected sources.
The optical timing dispersion is on the order of ns, longer than the detector jitter.
Cherenkov telescopes have poor throughput
at red wavelengths, where the SN signal
is strongest.
\added{For these reasons, we consider a Keck-like observatory designed to have a fine point-spread function and high red-IR throughput, while noting that it is likely more cost effective to obtain good optical quality and total collecting area with a larger number of smaller aperture telescopes.}

Supernova shape evolution can be modeled and measured through multiple observations over time. These repeated measurements yield multiple distance estimates, reducing systematic uncertainties inherent to supernova modeling.
In the analysis presented in this article
we assumed that the phase of the supernova
at the date of observation is known, say
as constrained by pre and postexplosion
photometric measurements by transient searches.  
However, this assumption becomes unnecessary when observations are taken on different dates. Two observations can constrain the phase directly, as this information is explicit in the SEDONA model and can be reconstructed empirically from ejecta velocities derived from single-phase TARDIS fits.

Geometric distances less sensitive to the details of the physical processes may be derived based on the angular expansion rates measured from intensity interferometry and the physical size of the photosphere that can be derived from the photospheric speed measurable from normal spectroscopy. Multiepoch intensity interferometry is needed to avoid the need for the exact time of explosion.  Distances can be calculated simply by dividing  the angular expansion rates of the photosphere by the corresponding photospheric velocities. However, spectroscopic models are still needed to select wavelength regions that can be robustly modeled by today's radiative transfer codes.

Core-collapse supernovae bright enough to produce an intensity interferometry signal are also expected to occur. While most core-collapse supernova subtypes are intrinsically more asymmetric than SNe Ia, type II-P supernovae are an exception. These events generally exhibit very low polarization during their plateau phase, and their modeling is relatively straightforward due to well-understood progenitor systems \cite{2018ApJ...861....1N, 2009MNRAS.395.1409S}. As such, they are valuable for distance determination using the expanding photosphere method 
\cite{1994ApJ...432...42S, 1996ApJ...466..911E, 2005A&A...439..671D, 2024arXiv241104968V}.

\section{Data Availability}
The data that support the findings of this article are openly available \cite{o_brien_2024_14511076}.

\begin{acknowledgments}
We would like to thank Jeff Hodgson,
Eddie Baron, Dan Kasen, Neal Dalal,
and Paul Stankus
for valuable discussions.
This work was supported by the U.S. Department of Energy (DOE), Office
of Science, Office of High Energy Physics, under Contract No. DE–AC02–05CH11231.
The SEDONA radiative transfer calculation used FASTER at TAMU through the allocation PHY240215 from the Advanced Cyberinfrastructure Coordination Ecosystem: Services and Support (ACCESS) program, which is supported by U.S. National Science Foundation Grants No.~\#2138259, No.~\#2138286, No.~\#2138307, No.~\#2137603, and No.~\#2138296.
\end{acknowledgments}

\bibliographystyle{apsrev4-2}
\bibliography{biblio}
\end{document}